# Analysis of ResearchGate, A Community Detection Approach

Mohammad Heydari*
School of Industrial and Systems Engineering
Tarbiat Modares Universityline
Tehran, Iran
m_heydari@modares.ac.ir

Babak Teimourpour
School of Industrial and Systems Engineering
Tarbiat Modares Universityline
Tehran, Iran
m_heydari@modares.ac.ir

***Abstract*— We are living in the data age. Communications over scientific networks creates new opportunities for researchers who aim to discover the hidden pattern in these huge repositories. This study utilizes network science to create collaboration network of Iranian Scientific Institutions. A modularity-based approach applied to find network communities. To reach a big picture of science production flow, analysis of the collaboration network is crucial. Our results demonstrated that geographic location closeness and ethnic attributes has important roles in academic collaboration network establishment. Besides, it shows that famous scientific centers in the capital city of Iran, Tehran has strong influence on the production flow of scientific activities. These academic papers are mostly viewed and downloaded from the United State of America, China, India, and Iran. The motivation of this research is that by discovering hidden communities in the network and finding the structure of intuitions communications, we can identify each scientific center research potential separately and clear mutual scientific fields. Therefore, an efficient strategic program can be design, develop and test to keep scientific institutions in progress path and navigate their research goals into a straight useful roadmap to identify and fill the unknown gaps.**

## I. INTRODUCTION

In the age of information, data known as the most important asset to the various types of large scale, medium and small enterprises around the world. Text data is ubiquitous and growing quickly. The web, blogs, emails, social networks provide millions of data in seconds and all of these resources can be a beneficial and powerful base for knowledge extraction and useful results. It's called the power of text analysis. Currently, social network platforms are under the scrutiny of many academic researchers and developers. The topic is widely utilized in various scientific, social, commercial and industrial aspects. Information systems can be mapped In Complex network form that contains linked nodes and relation. Social Network Analysis has a deep potential to acts a tremendous role in human activities around the world. By analysis of huge range of extracted data from various kinds of social networks, a number of useful hidden patterns emerges. Social Network Analysis is a type of structure analysis way expanding in many research fields which focuses on the related research and is mainly used to describe and measure the relationship and information individually. [3][4] Social Network Analysis has been proved to be successful in studies of scientific collaboration network. Interaction with other authors in social networks will increase the co-authorship, besides coauthorship increase citation. According to the ResearchGate scored the highest, 61.1 percent followed by Acadmia.Edu with 48.0 percent score ranked "above average" and Mendeley with 43.9 percent ranked "average". The success of RG has already enabled researchers to announce their ideas and share their publication free of charge to facilitate the remote access for the researchers all around the world. Based on Nature survey, RG has 2nd rank in Science and engineering topic and 4th in social science, arts, and humanities. It should be noted RG network became bigger compares to the Nature survey in 2014. To prove the mentioned claim, we looked up in Google Trend Service. The visualization of searched data based on keywords proves that the RG has the most interest to be visited among other similar networks. RG growth is increasing every day and now it has more than +15 million members all around the world and ranked 162 among all web pages in the world. Recently RG Score has become a major source of academic papers. The RG Online Scientific Network has a significant part of scholar's communication around the world. Based on a survey of Nature Journal, 48% of science and engineering researchers and 35% of social science, arts and humanities scholars visit RG regularly. The percent this social network is five more than Academia.Edu which is known as its nearest rival. In RG participants can utilize their own profile to share and represent their ultimate publications and projects. It must be noted that Graphical User Interface of RG is better than rivals. The search section of the site has been implemented intelligently and impressively cause by a single phrase search you can achieve various results about Researchers, Projects, Publications, Questions, Jobs, Institutions, and Departments that may search item relates to the mentioned filed. Since number of institutions haven't joined to, and they don't have official profiles, the score that system assigns to the scholars and institutions is still an argumentative problem. Based on RG Network, at the time of writing this paper, it's introducing a new trial metric to measure researchers score called Research Interest. Research Interest is how to measure scholarly interests in any research. This score is focused on research items and scientists' interactions with them, using concepts that are familiar to RG members. To provide an overview. This score is focused on research items and researcher interaction with them, using concepts that are familiar to RG members. To provide an overview of a researcher's body of work, RG also added a Total Research Interest score, which simply adds up the Research Interest scores from all an author's research items. When researchers read, recommend, or cite a research item, its Research Interest goes up. RG decided on a system for weighting the different forms of interaction based on a reader has a weighting of one, a full text read has a weighting of three, a recommendation has a weighting of five and a citation has a weighting of ten. A comparison of growth rate of RG against another famous academic portal is shown in Figure 1.

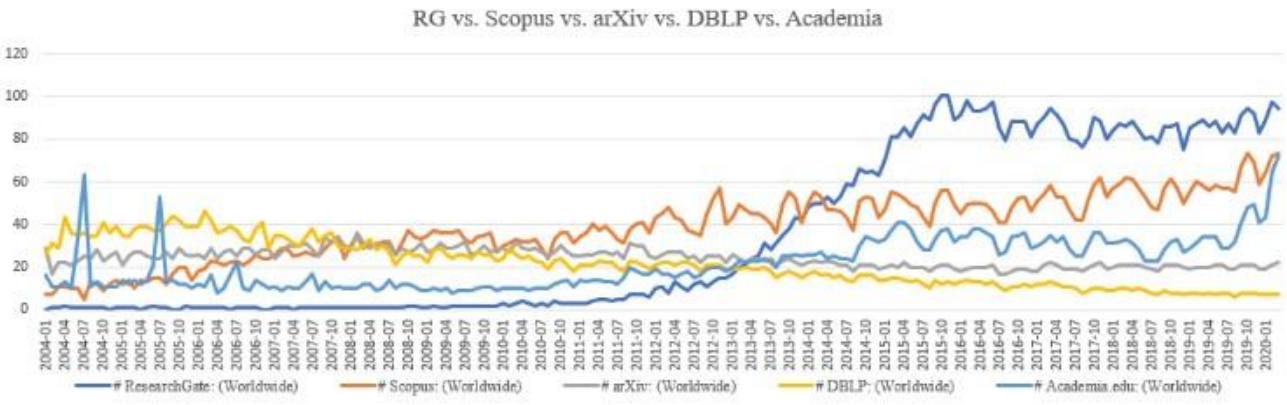

Figure 1 - Growth Rate of Interest in ResearchGate Network

## II. REALTED WORKS

Abbasi et al. extended a theory model based Social Network Analysis methods to investigate on collaboration network of Researchers by usage of famous Social Network Analysis methods (i.e., normalized degree, closeness, betweenness, eigenvector centralities and average ties strength and efficiency) for investigating consequence of social network on the Researchers efficiency on a certain topic (i.e., Complex networks). Abbasi also mentions that Researchers should work with lots of students as an alternative for the rest of wellknown Researchers. Manca et al. proposed a multi-level framework towards analysis of RG and Academia.Edu. His aim is two exemplify how these two online social networks are socio-technical systems that support researcher's knowledge activity sharing and professional learning. His proposed framework includes three layers: A. The "socioeconomic layer", B. the techno-cultural layer, C. the networked-scholar layer. Then he adopted described social networks to these three layers. Naderbeigi et al. made an investigation on mapping profile of research activities of faculty members of Sharif University of Technology in RG. They intend to test the correlation h-index between the RG and Web of Science and Scopus and Google Scholar. Then investigate Sharif University of Technology faculty members' top h cited research RG in WoS, Scopus, and GS. After all, investigate the Altimetric score of SUT faculty members' top h cited research RG with Altimetric Explorer. They already mention that more than %75 of Sharif University of Technology faculty member has already set their profile on RG. Besides that, the highest correlation of RG Score is with h-index and citation. Muscanell et al. examined usage and utility of RG by employing an online survey approach to target scientist who has an active RG profile. They find that most researchers who have an RG profile did not utilize usually. In the following, members did not perceive many benefits from RG profile and RG use was not considered to members work satisfaction or informational benefits but was considered to productivity and stress. Newman has already investigated in the scientific collaboration network by analyzing papers in computer science and physics in a 5-year period from 1995 to 1999. He constructed a collaboration graph based on a data extracted from for databases. The idea of collaboration patterns by using data extracted from scientific networks is not a new topic at all. However, to our knowledge, no similar collaboration network of all various types of Iranian scientific institutions has previously been attempted. Haiyan et al. worked on the structure of scientific collaboration network in Scientometric at the level of individuals by utilizing bibliographic data of published papers from 1978 to 2004. His novelty is the construction of a combined Social Network Analysis, Co-occurrence analysis, cluster analysis and frequency analysis of words to expose the collaborative center of the network, major collaborative fields of the network and various collaborative sub-networks and microstructure of the collaboration network. Šubelj et al. researched on convexity in the scientific collaboration network. They analyzed a particular dataset Slovenian researcher in computer science, physics, and other fields and identified convex skeletons in the collaboration network by eliminating weak links of them and provided frequency distributions of several data parameters of the skeleton and of remained graphs. Finally demonstrated that convex skeleton is a good abstraction of the original co-authorship network. To be clearer about convexity in a graph, it refers to an attribute of its subgraphs to comprise all shortest path between nodes of that subgraph. The scale values can be in [0, 1]. The biggest part of a graph that appears after the removal of the minimum number of an edge is called a convex skeleton which is fundamentally a tree of cliques. Perianes-Rodriguez et al. suggested a new fractional counting method to build bibliometric network which already compared with a traditional full counting method. In examination of university co-authorship network and journal bibliometric network, each one of two methods reveal distinct results. The first mentioned approach is known superior in various causes. Bihari et al. point out a key problem in research communities which is every author complete reputation of citation count on a particular paper, but in many cases their contribution is not the same at all. To eliminate this specific issue, they decide to utilize Poisson distribution to spread the contribution reputation in multi-authored paper and for examine research impacts of an independent, the eigenvector centrality has been utilized.

## III. RESEARCH METHODOLOGY

In the first step, ResearchGate selected as the target scientific network to study. necessary data from objective scientific institutions gathered straight completely for the main goal of the study. After the data preprocessing phase, the database prepared to be investigated. By the help of network science, we created the collaboration network graph. At the final step, Louvain algorithm applied to the network to identify hidden communities and the structure transparently.

### A. Data Crawling

Data is the heart of every data science project. The data gathered and collected from the profile page of each institution which has an active profile on RG until. Our research target covers all types of scientific institutions in Iran such as Public, Medical, Technical, and fundamental institutions. The main reason that medical, technical, and other types of institutions are considered in our study is covering various kind of theoretical communication among all major and academic fields at a comprehensive level. We don't aim to focus just on a particular research field. That's because different institutions type can be influential on results.

### B. Data Cleaning

Data cleaning is a fundamental method that initiate reshaping raw data into a readable style. After the data gathering phase,

the data cleaning task initiated to prepare them for graph construction from scratch. Many datasets are usually imperfect, incompatible with absence in certain behaviors, and is possibly to contain many missing values. Data preprocessing is a standard technique of eliminating such issues and preparing raw data for better processing.

**C.** Graph Construction

Since our approach is towards social network analysis field, in our study, nodes refer to a particular university and edges defines scientific collaboration among them. The relation model is shown in Figure 2.

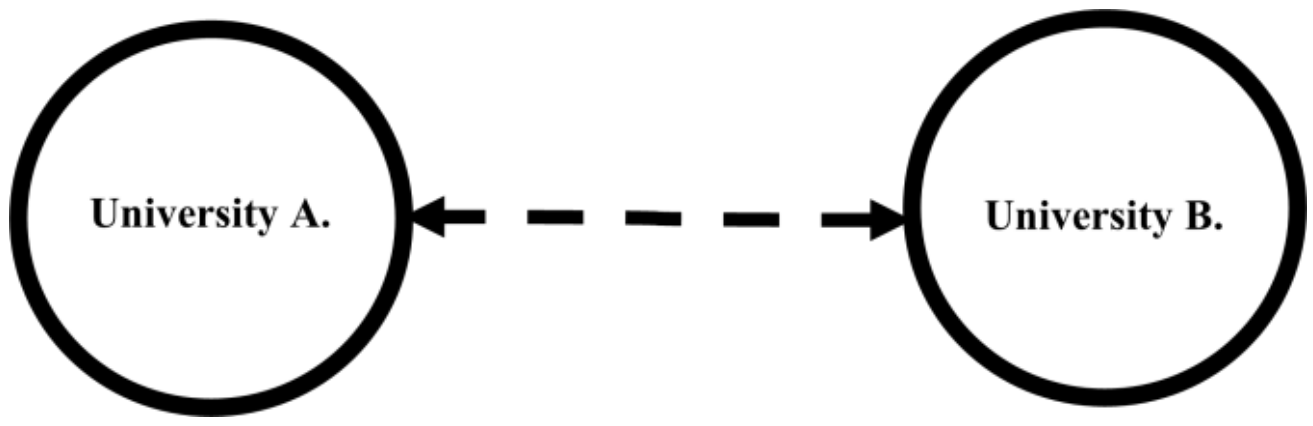

Figure 2 - Graph Relations

**D.** Network Centralities

Centrality methods role is significant in network science. In our study, we calculate four main centralities among them are Degree, Betweenness, Closeness and PageRank Centrality. The result is shown in Table 2.

**E.** Community Detection

The Louvain is a greedy optimization technique for community detection. It maximizes a modularity score for each community, where the modularity measures the quality of an allocation of nodes to communities by evaluating how much more densely connected the nodes in a community are, compared to how connected they would be in a random network. It is one of the fastest modularity-based algorithms. It also demonstrates a hierarchy of communities at different scales, which can be functional for understanding the global functioning of a network. The inspiration for this method of community detection is the optimization of modularity as the algorithm advances. Modularity is a scale value in the middle of [-1, 1]. It measures the density of edges inside communities to edges outside communities. Optimizing this value theoretically results in the best possible clustering of the nodes of a sample network, however going through all possible iterations of the nodes into groups is infeasible, so heuristic algorithms are utilized. In the Louvain algorithm, first small groups are found by optimizing modularity locally on all nodes, then each small group is organized into one node and the first step is repeated. The algorithm process is a heuristic method based on modularity optimization. The quality of the communities detected by this algorithm is better than previous methods as measured by modularity metric.

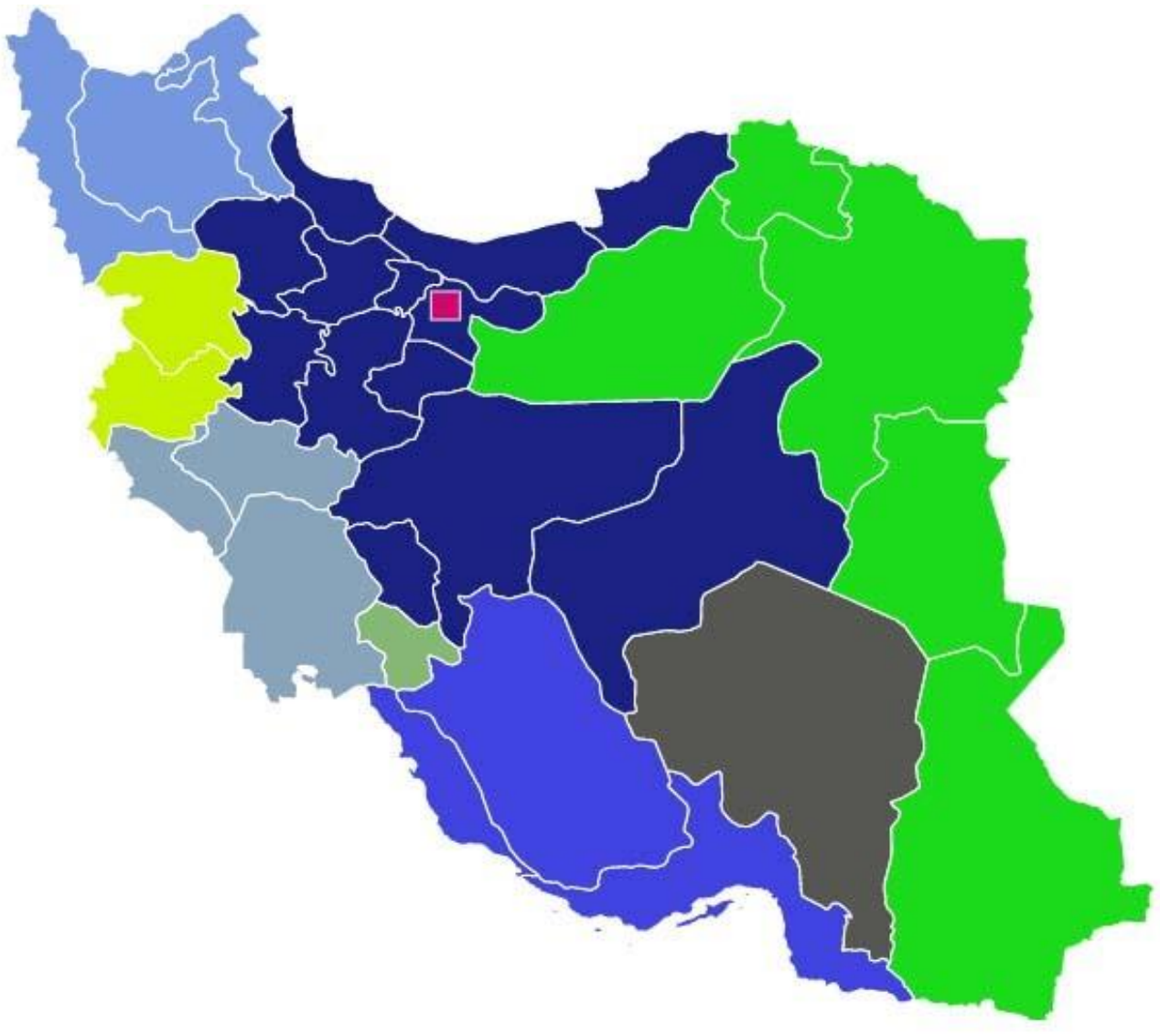

Figure 3 – Community Detection Findings

## IV. RESULTS

In the final section, we present the results of our study. with Average Degree 3.4973 it clears that each institute is at least connected to three other institution which can be conclude that our graph is not dense, and we encounter with a sparse space. Density value with 0.019 can be as proof for the claim. Based on degree centrality measure, the graph central node is University of Tehran. NetworkX graph library and Gephi visualization software employed to visualize the network in an explicit way. An interesting point is that institutions with the highest RG score are also ranked in highest position in Islamic World Citation and Webometrics ranking. The reason that Shahid Beheshti University of Medical (SBMU) and Iran University of Medical Science (IUMS) are ranked among top medical institutions is their strong connection with Tehran University of Medical Science (TUMS) which indicates the PageRank Phenomenon. It means SBMU and IUMS both recommended by a Strong Node which is (TUMS) that leading them both in Medical Section. Community structure is a common characteristic of social networks. nine communities identified by employing Louvain algorithm. The number of communities has not been fixed and obtained by the process of community detection. The method is like the earlier method that connect communities whose merges produces the largest community in modularity. To evaluate the quality of detected communities, it has been tested by Girvan-Newman clustering algorithm. the process indicates Maximum found modularity is 0.48. To demonstrate geographical presence of identified communities we already initiated them to the country map to have a big picture of the network structure as it is shown in the Figure 3. Each color refers to a particular community. The complete implementation of our study is available on the project's GitHub repository. Graph circular visualization, centralities correlations and graph Hubs are shown in Figure 4, Figure 5, and Figure 6 respectively. Finally, Centrality measurement calculation results are shown in Table 1.

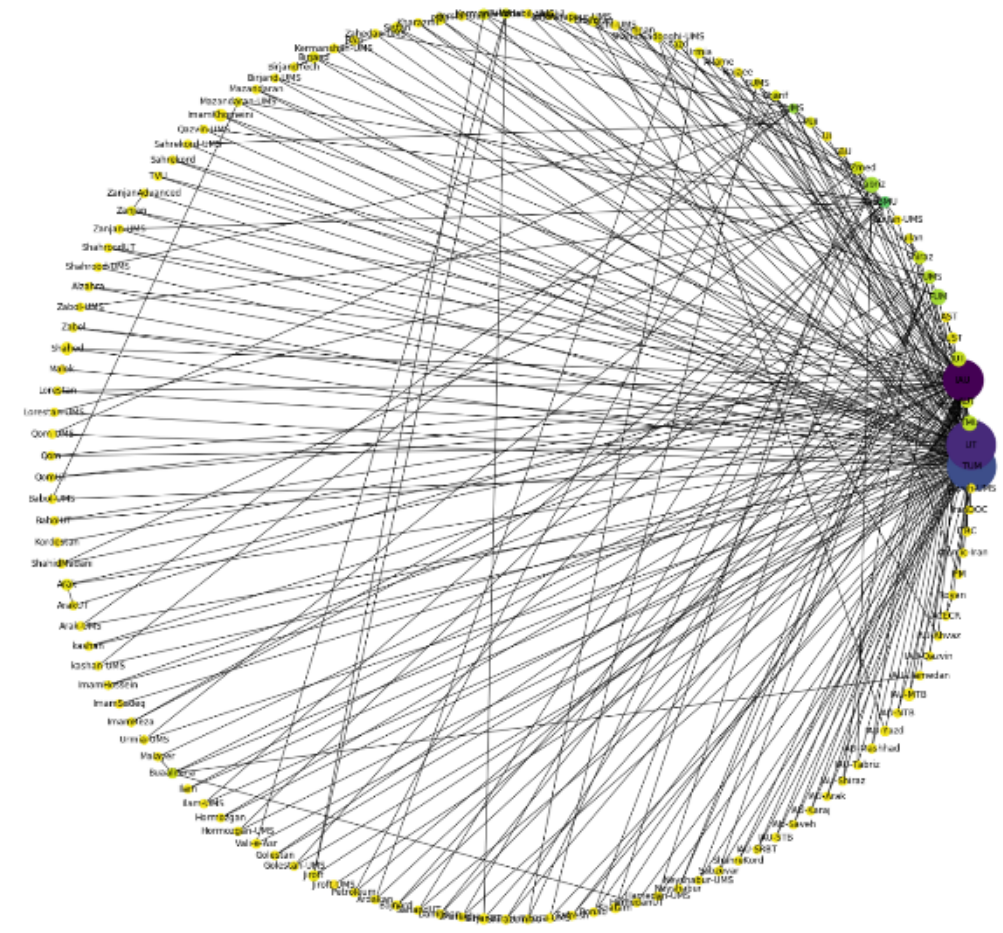

Figure 4 - Circular Visualization of the Network

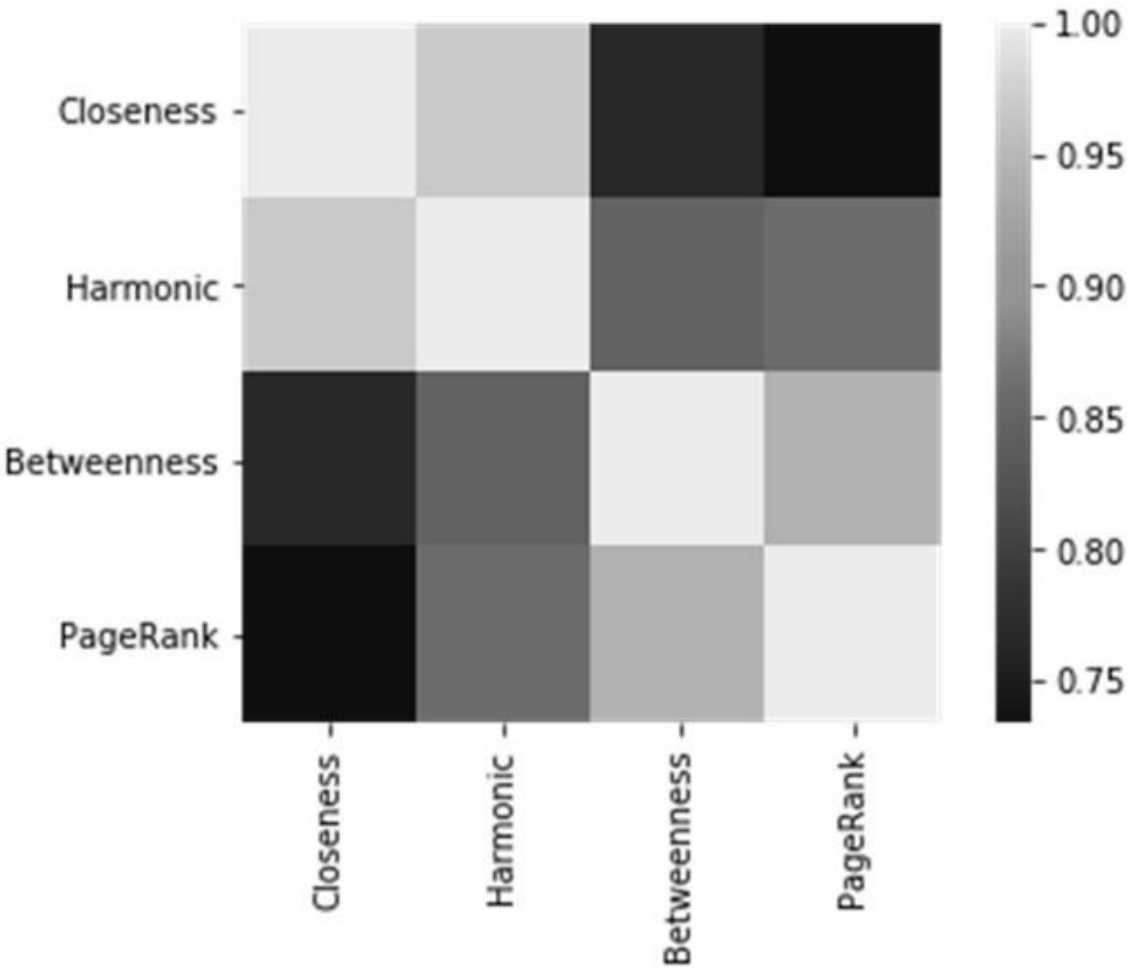

Figure 5 - Centralities Correlation of the Network

## V. CONCLUSION

In this study we utilized network science techniques to create, analyze and visualize collaboration network of Iranian scientific institutions for the first time. The contribution of our research is discovering structure of communities within the network. Our results demonstrated that geographic location closeness and ethnic attributes has important roles in academic interactions establishment. Besides, it shows that famous scientific centers in the capital city of Iran, Tehran, has strong influence on the production flow of scientific activities. The output of this study can be beneficial for Science and Technology Policy Organization who aims to design and initiates comprehensive strategies to gain more profit of conducted research and development in various knowledge-based fields. The most important benefit of community structure awareness of our network is identifying each scientific institute's research potential separately and recognize hot topic fields.

Appendices. (Heydari & Teimourpour, 2020).

Table 1 - Centralities Calculation Results to Identify Key Nodes on The Network

| ***Nodes*** | ***Degree*** | ***Nodes*** | ***Page Rank*** | ***Nodes*** | ***Betweenness*** | ***Nodes*** | ***Closeness*** | ***Nodes*** | ***Harmonic*** |
|---|---|---|---|---|---|---|---|---|---|
| IAU | 0.36263 | IAU | 0.109805 | UT | 0.483103 | UT | 0.628019 | UT | 0.703846 |
| UT | 0.346153 | UT | 0.095966 | TUMS | 0.473596 | TUMS | 0.543933 | IAU | 0.678205 |
| TUMS | 0.280219 | TUMS | 0.084518 | IAU | 0.321358 | IAU | 0.539419 | TUMS | 0.644872 |
| SBMU | 0.104395 | SBMU | 0.029387 | FUM | 0.032998 | TMU | 0.460993 | TMU | 0.489744 |
| TMU | 0.065934 | IUMS | 0.018656 | Tabriz | 0.031145 | IUT | 0.431894 | IUT | 0.462821 |

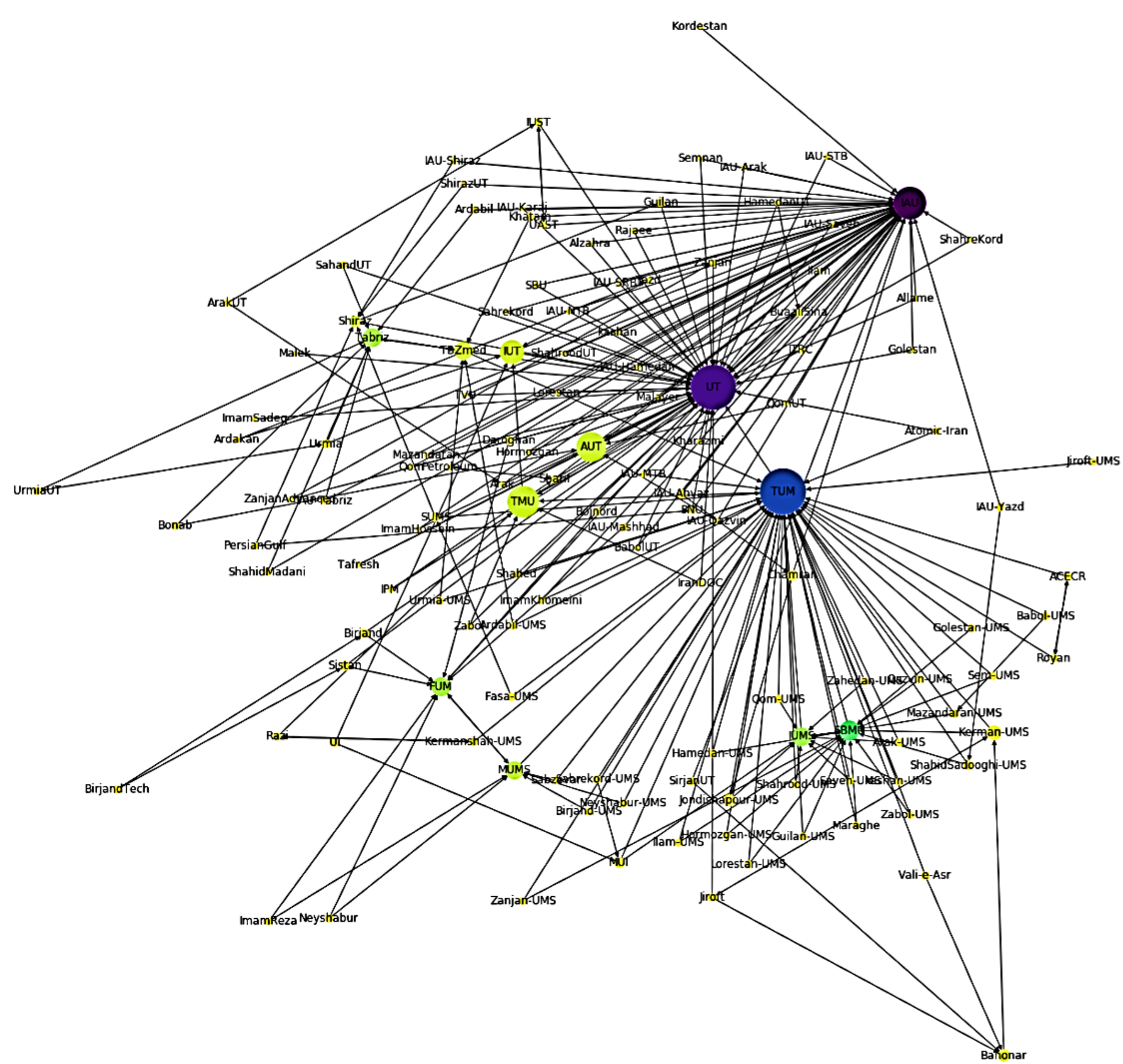

Figure 6 - Graph Hubs are Highlighted.